\documentclass[11pt]{article}

\usepackage{url} 
\usepackage{graphicx,amssymb,amsmath,amssymb,textcomp,url,booktabs,lineno}
\usepackage[left=1.5in,right=1.5in,bottom=1.5in,top=1.5in]{geometry}
\usepackage{natbib}

\title{Modelling hippocampal neurogenesis across the lifespan in seven species}

\author{Stanley E. Lazic  \\
  Bioinformatics and Exploratory Data Analysis, F. Hoffmann-La Roche \\
  4070 Basel, Switzerland \\
  \url{ stan.lazic@cantab.net}
	}

\date{}

\begin{document}

\maketitle

\begin{abstract}
The aim of this study was to estimate the number of new cells and neurons added to the dentate gyrus across the lifespan, and to compare the rate of age-associated decline in neurogenesis across species. Data from mice (\textit{Mus musculus}), rats (\textit{Rattus norvegicus}), lesser hedgehog tenrecs (\textit{Echinops telfairi}) macaques (\textit{Macaca mulatta}), marmosets (\textit{Callithrix jacchus}), tree shrews (\textit{Tupaia belangeri}), and humans (\textit{Homo sapiens}) were extracted from twenty one data sets published in fourteen different papers. ANOVA, exponential, Weibull, and power models were fit to the data to determine which best described the relationship between age and neurogenesis. Exponential models provided a suitable fit and were used to estimate the relevant parameters. The rate of decrease of neurogenesis correlated with species longevity r = 0.769, p = 0.043), but not body mass or basal metabolic rate. Of all the cells added postnatally to the mouse dentate gyrus, only 8.5\% (95\% CI = 1.0\% to 14.7\%) of these will be added after middle age. In addition, only 5.7\% (95\% CI = 0.7\% to 9.9\%) of the existing cell population turns over from middle age onwards. Thus, relatively few new cells are added for much of an animal's life, and only a proportion of these will mature into functional neurons. 
\end{abstract}

\section{Introduction}

Levels of hippocampal neurogenesis have been shown to correlate with certain learning and memory tasks \citep{Shors2001} as well as performance on measures of depression in rodents \citep{Santarelli2003}. New neurons generated in the hippocampus of adult animals can functionally integrate into the neural circuitry \citep{vanPraag2002,Kee2007} and have different electrophysiological properties compared to existing cells \citep{Schmidt2004}, suggesting that they may play a unique role in information processing. Since levels of neurogenesis can only be determined postmortem (although a novel NMR approach may circumvent this, \citealt{Manganas2007}), it is difficult to estimate the number of new neurons that functionally integrate into the dentate gyrus (DG) throughout the life an organism. The size of the DG does not increase with age and the number of granule cells remains fairly constant \citep{BenAbdallah2010}. A commonly cited number of new neurons is 6\% of the total number of cells in the granule cell layer (GCL), or 30-60\% of the afferent and efferent population of neurons, based on an early paper by \citet{Cameron2001}. However, this was based on a single time point in young adult rats and cannot be extrapolated to older animals, and it is well known that age has one of the largest effects on neurogenesis \citep[and see Table 2 in the Supplementary Material]{Kuhn1996}.

Measuring the number of new neurons is difficult because markers of proliferation  measure all dividing cells and are not neuron specific. Labelling cells with BrdU and (after a suitable delay to allow the cells to mature) a neuronal marker is problematic because of the complex dynamics of BrdU-labelled cells: they continue to divide for several days after initial labelling \citep{Hayes2002}, thus making it difficult to determine the number of cells that were undergoing cell division at time of BrdU injection. In addition, the number of labelled cells is proportional to the dose of BrdU \citep{Cameron2001}. Using retroviral vectors is another method, but this technique may not label all dividing cells (see \citet{Breunig2007} for a review). Markers of immature neurons such as doublecortin (DCX) and neurogenic differentiation factor (NeuroD) can also be used to determine the number of new neurons \citep{Couillard-Despres2005,Bohlen2007}, but these markers are expressed for several weeks, and therefore it is difficult to estimate the number of new neurons at a single time point, such as the number of new neurons added in a single day. This makes it difficult to estimate the total number of cells added during the life of an animal, as cell cycle kinetics need to be taken into account.

The literature was therefore searched for publications that examined proliferation or neurogenesis in the hippocampus at multiple time points. Data were extracted from these papers and used to estimate the relationship between neurogenesis and age, and to derive some important quantities, such as the total number of proliferating cells added during the lifespan, and the number of new neurons added during various life stages. In addition, the majority of studies used rats and mice, and a quantitative comparison across species has not been conducted, which might elucidate evolutionary and phylogenetic aspects of neurogenesis.

Data from \citet{BenAbdallah2010} were used to determine the functional form of the relationship between neurogenesis and age (i.e. which model was best). Once this was determined, a model of the same functional form was fit to all of the other data sets, and the parameters estimated. The rate of decrease in proliferation and neurogenesis was the main parameter of interest and was correlated with species longevity, body mass, and basal metabolic rate (BMR). Next, estimates were made of the total number of cells added to the DG, the number of new neurons added, and how the addition of new cells was distributed across different life stages. Data from \citet{BenAbdallah2010} was used for these estimates.

\section{Materials and methods}

\subsection{Data}
Data were obtained from papers indexed in PubMed with the following search criteria: ``neurogenesis AND (hippocamp* OR dentate) AND (age OR aging)'', up to and including papers from 28 February 2010. In addition, reference lists from these papers were examined for other potential studies. Only those papers that contained at least four time points were used, and data were accurately extracted from the graphs using g3data software (\url{www.frantz.fi/software/g3data.php}). Analysis was conducted with R (version 2.10.1; \citealt{Ihaka1996,R2010}).

\subsection{Model fitting and parameter estimation}

Only data from \citet{BenAbdallah2010} were used to compare the various models. This study was the only suitable one because the raw data were presented rather than summary statistics (e.g. means and error bars), observations were made at seven time points, a reasonable estimate of the intercept could be obtained because juvenile animals were included, there were approximately five replicates at each time point, Ki67 was used as a marker of proliferation (BrdU kinetics would have added greater complexity), modern stereological methods were used, and results were reported as the absolute number of labelled cells. Several empirical models were fit to the data, including ANOVA, exponential, biexponential, Weibull, and power. The biexponential model did not converge \citep[likely because it was not a suitable model, and the difficulty of estimating parameters from sums of exponentials;][]{Bates1988}, and was not considered further. Models were compared using an information-theoretic approach described in detail in \citet{Burnham2002} and \citet{Anderson2008}.

In the models below, $t$ is the age of the animals (in months), and $N(t)$ is the number of labelled cells at age $t$.

The ANOVA model has the form 

\begin{equation}
\mathrm{ANOVA:} \quad N(t) = \mu + \theta_t \label{equ:anova}
\end{equation}

\noindent where $\mu$ is the overall mean and $\theta$ is the difference between the overall mean and the mean of the animals at age $t$. This is the most common model used to analyse the relationship between age and neurogenesis, and it has the advantage of being the most flexible, but at the cost of greater complexity, less precise estimates, more difficult to interpret, and leads to unnecessary post-hoc tests \citep{Lazic2008}. Nevertheless, it can be used as a baseline model to compare the others against. The exponential model has the form

\begin{equation}
\mathrm{Exponential:} \quad N(t) = \beta\, e^{-\alpha\,t} + \gamma \label{equ:exp}
\end{equation}

\noindent where $\alpha$ is the slope (how quickly the number of labelled cells decrease over time), $\gamma$ is the lower asymptote (which allows neurogenesis to decrease to a non-zero level), and $\beta$ is the ``drop''---the difference between the number of cells at birth ($t$=0) and the lower asymptote. If $\gamma=0$ then $\beta$ is the $y$-intercept (the number of labelled cells at birth). The value of the slope can also be expressed more intuitively as a half-life ($t_{1/2} = \frac{\mathrm{ln}\,2}{\alpha}$), which is the length of time (in months) for the number of labelled cells to decrease by half.  At least two previous studies \citep{Simon2005,BenAbdallah2010} recognised that the relationship between proliferating cells and age can be described by an exponential model. The Weibull model has the form

\begin{equation}
\mathrm{Weibull:} \quad N(t) = \beta\, e^{-\alpha\,t\,^\delta} + \gamma \label{equ:weib}
\end{equation}

\noindent where $\alpha$, $\beta$, and $\gamma$ have the same interpretation as in the exponential model, and the additional parameter $\delta$ allows one to model a changing rate of proliferation or neurogenesis over time. If $\delta = 1$, then the Weibull model reduces to the exponential model, and this means that the rate of decrease is constant over time. A value of $\delta > 1$ means that as animals age, levels of proliferation/neurogenesis decrease more quickly, while a value of $\delta < 1$ means that the rate of decrease slows down over time. The power model has the form

\begin{equation}
\mathrm{Power:} \quad N(t) = \beta\, t^{-\alpha} + \gamma  \label{equ:power}
\end{equation}

\noindent where $\alpha$, $\beta$, and $\gamma$ have the same interpretation as above, although $\alpha$ cannot be converted to a half-life using the above equation. \citet{Knoth2010} fit a straight line to their data when plotted on a log-log graph, and while they did not discuss power functions, these are commonly plotted on such a graph and suggests that this might be another suitable model.

For most data sets, younger animals not only had higher mean values of labelled cells, but also greater variability compared to older animals. If this was the case, and the raw data were available, the errors were modelled as function of the fitted values ($\mathrm{Var}(\epsilon_{i}) \sim \mathrm{N}(0,\sigma^2 \times \mathrm{fitted}_i^{2 \delta})$). If only summary data were available, the means were weighted proportional to their precision (reciprocal of the standard error).

\subsection{Clustering and regression}
Hierarchical clustering (Euclidean distance with complete linkage) was used to visualise the rates of decrease in proliferation and neurogenesis between the species. If more than one data set was available for a species, the mean was calculated and used in the clustering.

Data on species body mass (grams) and basal metabolic rate (mL\,O$_2$/hr) were obtained from \citet{Sieg2009}, and maximum lifespan data (years) were obtained from The Animal Ageing \& Longevity Database (\url{http://genomics.senescence.info/species/}). These variables were used in a weighted regression analysis, with the half-life values for proliferation and neurogenesis used as the outcome variable. Half-life values were weighted according to their estimated precision.

\subsection{Number of cells added during various life stages}

To calculate the number of cells added to the DG, the unit of time was converted from months to days. This is convenient since the cell cycle is approximately 24 hours and Ki67 is expressed for a similar length of time \citep{Cameron2001,Mandyam2007}. However, the number of Ki67$^+$ cells counted is only an approximate value, and there are two adjustments that need to be made to the number of counted Ki67$^+$. First, the number of cells that divided at day $t$ is required, and not those that divided the previous day but are still expressing Ki67. \citet{Mandyam2007} estimated that approximately 33\% of cells labelled with BrdU 24 hours previously, still expressed Ki67. In other words, only 67\% of cells viewed on a section divided at day $t$, and the other 33\% divided the previous day. Second, the number of cells counted is not necessarily the number of cells added. Since Ki67 is expressed throughout the cell cycle (except G$_0$), a proliferating cell that is pre-mitotic will be counted as one labelled cell, whereas if it is post-mitotic (but still expressing Ki67), then two labelled cells would be counted. It is important to estimate the number of cells added, not just the number counted.  For example, if five cells divide, there will be a total of ten cells, but only five additional cells. It is therefore assumed that only 75\% of the number of counted cells are additional cells (see the Supplementary Material for the derivation of this value). Whether a recently divided cell is counted as one or two cells will depend partly on the magnification used, as a higher resolution will make it easier to distinguish between two closely juxtaposed cells. Therefore, taking these two considerations into account, if $n$ labelled Ki67 cells are observed, there are only approximately $n \times 0.67 \times 0.75$ new cells that have been added to the dentate gyrus. One assumption that has been made is that if a cell divides, then it expresses Ki67, and is thus counted. 

By integrating (calculating the area under the curve) of the exponential model (Eq. \ref{equ:exp}), the number of cells that have been added during any time interval can be estimated. In this model, the $\gamma$ parameter was not significantly different from zero (p = 0.197), suggesting that neurogenesis may stop completely in the long run. If this is true, then inclusion of the $\gamma$ term would bias the results upwards. However, $\gamma$ may not have been significant because the power was low (Type II error), and then removing this term from the model would bias the results downwards. Fortunately, one is not forced to chose one model over another based on an arbitrary p-value criterion. Estimates were therefore made from both models (with and without the $\gamma$ term) and averaged, but were weighted according to how well each model fit the data. This is similar to the previous model comparisons (ANOVA vs. exponential vs. Weibull vs. power). The weight was 0.60 for the model with no $\gamma$ term, and 0.40 for the model with the $\gamma$, thus the estimates are pulled closer to the model with no $\gamma$ term. 95\% confidence intervals for the estimates were obtained using a nonparametric bootstrap \citep{Davidson1997}.

\section{Results}

\subsection{Establishing the form of the relationship between age and proliferation/neurogenesis}

A visual examination of the graphs from the various studies (see Supplementary Material), suggests that the number of labelled cells declines exponentially with age. However, only one study fit an exponential model to the data \citep{BenAbdallah2010}, and one other study first log-transformed the data and then used a linear regression analysis \citep{Simon2005}. Linearising nonlinear relationships was a useful technique to make analysis easier (or even possible) before computers came into routine use, but such transformations are generally unnecessary today, as the error structure can be distorted and parameter estimates may be different compared to the corresponding nonlinear model \citep{Bates1988,Motulsky2004}. 

In order to determine the accuracy of the data extraction, the same analysis as \citet{BenAbdallah2010} was performed. The approximation error was less than 1\% for both parameters, with a calculated slope of -0.473 and $y$-intercept of 10470 (original: -0.475 and 10510).

The number of Ki67 and DCX cells from \citet{BenAbdallah2010} are plotted in Figure 1, along with the ANOVA, exponential, Weibull, and power models.  All of the models fit the data reasonably well, and the exponential model was the most suitable for both the Ki67 and DCX data. The complete results of the model comparisons are presented in the Supplementary Material.

\subsection{Estimating parameters from published studies}

An exponential model was therefore fit to twenty-one data sets from fourteen separate studies. The rate of decrease (slope or half-life) in proliferation and neurogenesis was the main parameter of interest, and these are displayed in Figure 2. The similarities between species can be better appreciated in Figure 3A and 3B, where it can be seen that the rodents form one cluster, while the other animals are in a second cluster. The numeric results and graphs of the fitted models are presented in the Supplementary Material, where it can be seen that such a model fits the data well across all studies.

Unfortunately, many of these studies reported values as the number of labelled cells per area or volume. This not only makes it impossible to extract parameters to determine the total number of cells, but the reported value is a function of both the number of cells (the value of interest) and the size of the region of interest. This means that without further information on the volume of the structure, it is not possible to determine whether changes at different ages reflect changes in cell number or the size of the structure. This can lead to a ``reference trap'' \citep{Oorschot1994,Mayhew2003}, and is the reason why modern stereological methods report values as the total number of cells in the structure of interest and not as density estimates \citep{Mouton2002,Baddeley2005}. A similar problem can arise in studies using \textit{in situ} hybridisation to measure gene expression  when the mean grey level of autoradiographic images is multiplied by the area of the structure, and an ``integrated'' value is reported \citep{Lazic2009}.

\subsection{Relationship with longevity, body mass, and BMR}

Proliferation half-life (Ki67$^+$) values were not associated with longevity, body mass, or BMR. As can be seen in Figure 3C, the half-life estimates were quite variable for the mouse studies. Neurogenesis half-life (DCX$^+$ cells) was associated with maximum lifespan (r = 0.769, p = 0.043; Figure 3C). It is also clear that the half-life values for the lesser hedgehog tenrecs were quite different from the other species. This is likely due to inaccuracy in estimating the half-life, since the youngest animals in this study were already two years old. If data for the hedgehog tenrecs are excluded, then the strength of the relationship between neurogenesis half-life and maximum lifespan increases substantially (r = 0.992, p $<$0.0001). This relationship is perhaps not surprising, and suggests that the number of new cells added is scaled according to lifespan, such that all animals get a similar proportion of new cells across the various life stages (i.e. short-lived species do not die with massive amounts of neurogenesis, while long-lived species have no neurogenesis for 90\% of their life).

\subsection{Number of cells added during various life stages}

The initial intention was to pool information across studies, but the different experimental methods and markers used, definition of the region of interest, and ways of expressing the final results meant that the data could not be combined in any meaningful way. Data from \citet{BenAbdallah2010} were used again for the reasons mentioned above, and in addition, they also reported the total number of cells in the DG. 

Developmental ages of mice and corresponding human ages are taken from \citet{Harrison2010}  \url{http://research.jax.org/faculty/harrison/ger1vLifespan1.html}. Given an average lifespan of 28 months for mice, it is possible to determine the total number of cells that have been added during the life of an average mouse by integrating Equation \ref{equ:exp} between 0 and 28 months. This gives a value of 368,500 cells. If we take the total number of cells in the dentate gyrus to be approximately 550,000 \citep{BenAbdallah2010}, then this represents 67\% (95\% CI = 60.5\% to 73.4\%) of the adult dentate gyrus cell population. This is the number of new cells that have been added to the GCL postnatally, and represents an upper limit to the number of new neurons that can be added, since only a certain proportion will differentiate into neurons. If we take 3 months as the age at which a mouse becomes an adult (corresponding to approximately 20 years in humans; \citealt{Harrison2010}), then the number of new cells added in adulthood is 114,300, or 20.8\% (95\% CI = 15.7\% to 25.6\%) of the mature DG population. This corresponds to 31.0\% (95\% CI = 23.6\% to 38.1\%) of all the cells that will be added postnatally (Figure 4). From middle age (10 months) onwards, only about 31,200 new cells will be added. This corresponds to 8.5\% (95\% = 1.0\% to 14.7\%) of all the cells that will be added postnatally, representing  5.7\% (95\% CI = 0.7\% to 9.9\%) of the mature population; in other words, only 5.7\% of the mature DG population turns over after middle age. The interesting result is that there is little addition of new cells in adults, particularly from middle age onwards.  Based on these numbers, it is straightforward to calculate the number of new neurons; for example, if we assume that 50\% of these cells mature into neurons (this number may vary according to the experimental manipulation, and perhaps species and other factors), then the number of cells is simply divided by two. Therefore, the number of new neurons added postnatally as a percent of the total number of cells in the DG is 10.4\%, which is close to the 6\% value calculated by \citet{Cameron2001} for rats, who did not take into account changes in proliferation across the lifespan, but were fortunate to have selected an ``average'' value. Here, Cameron and McKay's results are extended to show that the average is not representative, since the number of new neurons varies several orders of magnitude from birth to old age. Indeed, a key finding is that for the majority of an animal's life (middle age to death---10 to 28 months---is 64\% of a lifetime in mice) the number of new cells added is only 5.7\% of the existing population, and the number of new neurons will be even less.

\section{Discussion}

These results have implications for the role of adult neurogenesis on cognitive and affective behaviour; in particular, the finding that relatively few cells are added during the majority of an animal's life. While the dramatic decrease in neurogenesis in old age has been known for many years \citep{Kuhn1996}, it has perhaps not been appreciated how little neurogenesis actually occurs during much of adulthood. The contribution that neurogenesis makes to behaviour is difficult to determine because experimental methods that alter neurogenesis  (e.g. physical activity, stress, irradiation) have numerous off-target effects, and these are usually not taken into account (\citealt{Lazic2010a}, and see open peer commentaries: \citealt{Shamy2010,Shetty2010,Stranahan2010,Yee2010}). This is analogous to determining the relationship between occupational exposure to asbestos and the risk of lung cancer, while ignoring smoking status (a major known risk factor that may be correlated with occupation) and age (which is a risk factor for many types of cancers). Of note, when other relevant factors are taken into account, the relationship between neurogenesis and behaviour is no longer present \citep{Merrill2003,Bizon2004,Castro2010,Lazic2010a}. There are also numerous studies that do not find a relationship between neurogenesis and behaviour \citep[and references therein]{Jaholkowski2009} and one wonders whether neurogenesis might be an epiphenomenon, which does not have a causative role in behaviour, but which correlates with many processes that do. The shear number and variety of factors that affect neurogenesis, and to such a large extent, might suggest that neurogenesis is not physiologically important, since key physiological parameters tend to be kept within a narrow range (e.g. body temperature, pH, blood pressure, etc.). Perhaps hippocampal neurogenesis is best viewed as a general marker of ``brain health''; it is high in young, physically active, and cognitively enriched animals, with adequate levels of neurotransmitters, growth factors, etc., and low in aged, stressed, inflamed, diseased, and environmentally impoverished animals. If this is true, then attempts to increase hippocampal neurogenesis \textit{per se} for therapeutic ends would be a misguided strategy. However, given that neurogenesis often correlates with behavioural tests, levels of neurogenesis might still serve as a biomarker to screen for potentially useful interventions.

From these results it can be seen that levels of adult neurogenesis have been overestimated because (1) younger animals are typically used for experiments, and for much of an animal's life, neurogenesis is at a much lower level, (2) markers such as DCX are expressed for several weeks, and therefore many labelled cells are present on histological sections, giving the appearance of a large phenomenon, (3) cells continue to divide after labelling with BrdU, and thus protocols with a lag of more than 24 hours after injection will have more labelled cells than the number that divided at the time of the injection. In addition, some BrdU protocols have multiple injections over several days.

It should be noted that in the experimental data from \citet{BenAbdallah2010} the age of the mice ranged from one to nine months, and therefore the calculation of total proliferation over an average lifespan entailed extrapolating one month into the past and 19 months into the future. While this introduces some uncertainly into the results, the majority of proliferative activity occurred before nine months, and so uncertainties in the estimate of proliferative activity after this time would not have a large influence on the estimates. In addition, the model averaging approach used provides a certain robustness to the estimates because they are not dependent on a single model \citep{Burnham2002,Anderson2008}.

In this study, estimates were made from mice and extrapolated to humans, as there was no human study with comparable data. \citet{Knoth2010} provided the richest data set, but did not express the results as the total number of cells (but rather as the number/mm$^2$), which would have required access to the full hippocampus and which is difficult to obtain for human samples. Nevertheless, the exponential model also fit the human data well, suggesting that the results calculated from the mouse data would translate to humans (the usual caveats of extrapolating from animals to humans apply). As an interesting comparison, a 30 year-old human (around the time many people finish their PhD, and corresponding to 6 months in a mouse), will have already had about 86\% (95\% CI = 79\% to 94\%) of the new cells that they will ever get. There is a 75\% decrease in the number of new cells added per day between 20 and 30 years of age (3 to 6 months in a mouse), and despite such a large reduction, 30 year-olds do not seem to have gross impairments in spatial navigation or pattern separation compared to 20 year-olds. 

Wu et al. stated that ``major declines of BrdU$^+$, DCX$^+$, and BrdU$^+$/DCX$^+$ cell numbers occur before middle age, which represents a critical turning period of adult hippocampal neurogenesis'' \citep{Wu2008}. However there is no evidence for a ``critical turning point'' for neurogenesis, either in their study or in the other time-course studies examined---instead, there is a smooth decrease over time. In a recent review,  \citet{Amrein2009} suggested neurogenesis peaks at puberty and that this might have something to do with the transition from juvenile to adult behaviour. Similarly, there is little evidence for neurogenesis peaking postnatally (Amrein \& Lipp's statement was also taking into consideration the number of apoptotic cells).

Given the large change in neurogenesis across the lifespan, unqualified statements of thousands of new neurons being added each day to the hippocampus of the adult brain \citep[e.g.][]{Shors2008,Aimone2009} can be misleading, since for the majority of adult life, the number is far less.  To understand the role of neurogenesis in emotional or cognitive processing (if any), it needs to be examined across the lifespan \citep{Coleman2004}, since for the majority of a mammal's life, neurogenesis occurs at relatively low, and perhaps negligible levels.

\bibliographystyle{apa}
\bibliography{modbib}

\clearpage
\begin{figure}
\centering
\includegraphics[scale=0.5]{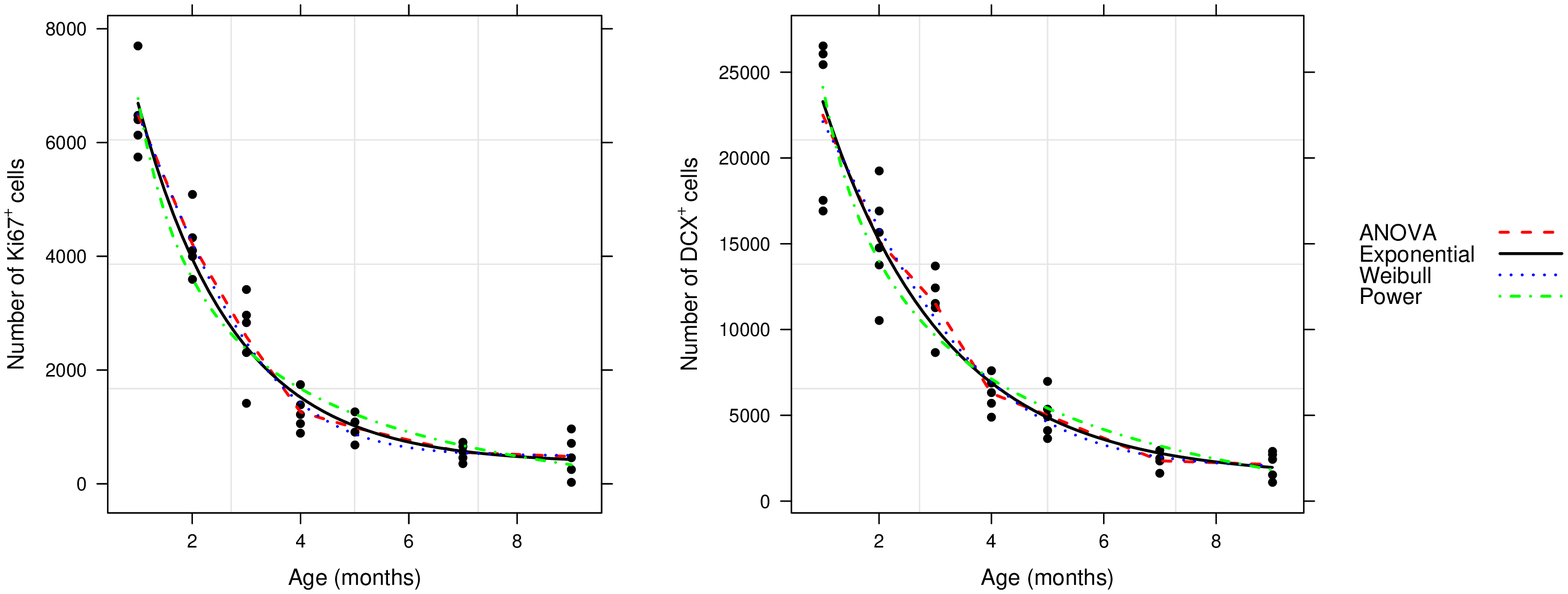}
  \caption{Model comparisons. All four models fit the data well and the exponential model was the most suitable. The red dashed lines go through the mean of each time point (ANOVA model). The power model performed substantially worse than the others, for both Ki67 and DCX data. Data were extracted from \protect{\citet{BenAbdallah2010}}, and numeric results can be found in the Supplementary Material.}
\label{fig:expfit}
\end{figure}

\clearpage
\begin{figure}
\centering
\includegraphics[scale=0.5]{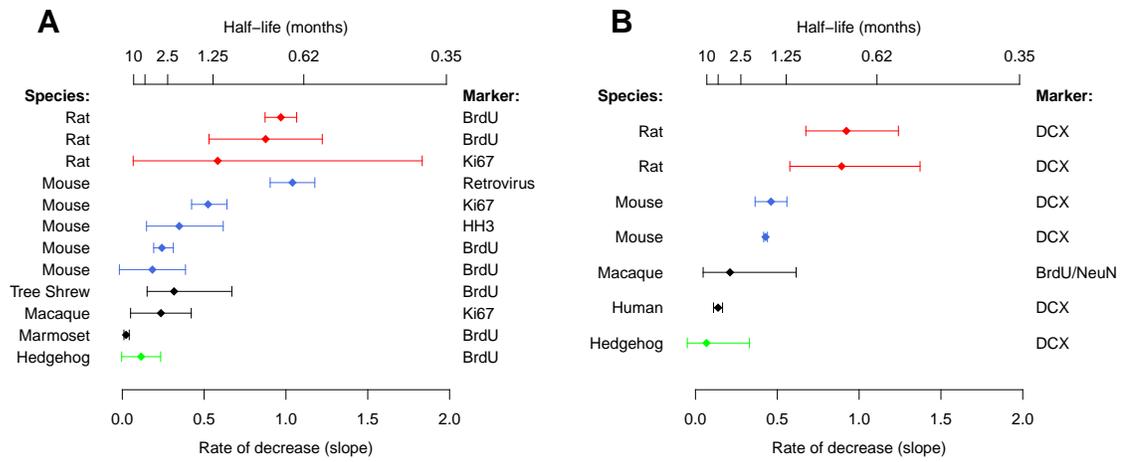}
  \caption{Rate of decrease of proliferation (A) and new neurons (B) across the lifespan in seven species. Error bars are 95\% confidence intervals. Generally, longer-lived species had a slower rate of decline in proliferation and neurogenesis.}
\label{fig:halflife}
\end{figure}

\clearpage
\begin{figure}
\centering
\includegraphics[scale=0.75]{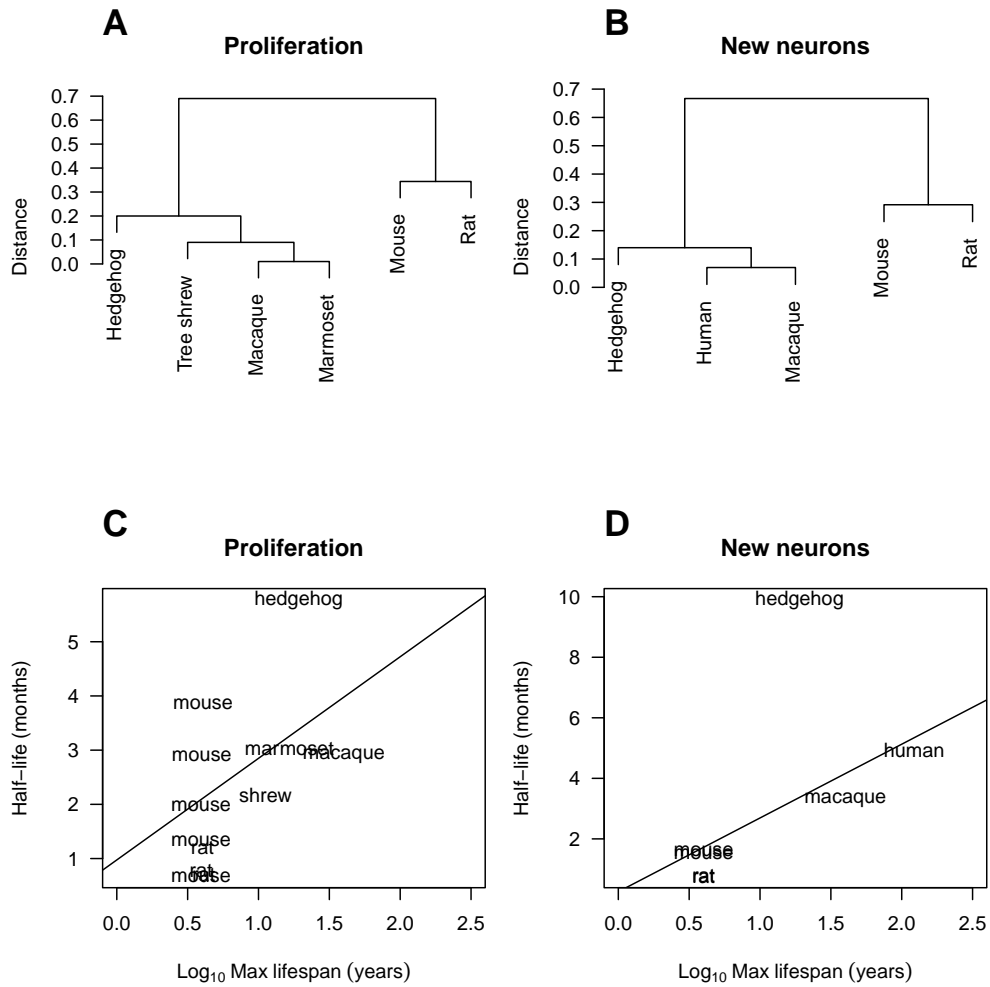}
 \caption{Clustering of species by rate of decrease in proliferation (A) and neurogenesis (B). Mice and rats formed one cluster, while primates and other species formed another. The association between lifespan and proliferation was not significant (r = 0.470, p = 0.123; C), but the association with neurogenesis was (r = 0.769, p = 0.043; D). Solid lines are regression lines.}
\label{fig:clust_mass}
\end{figure}

\clearpage
\begin{figure}
\centering
\includegraphics[scale=0.75]{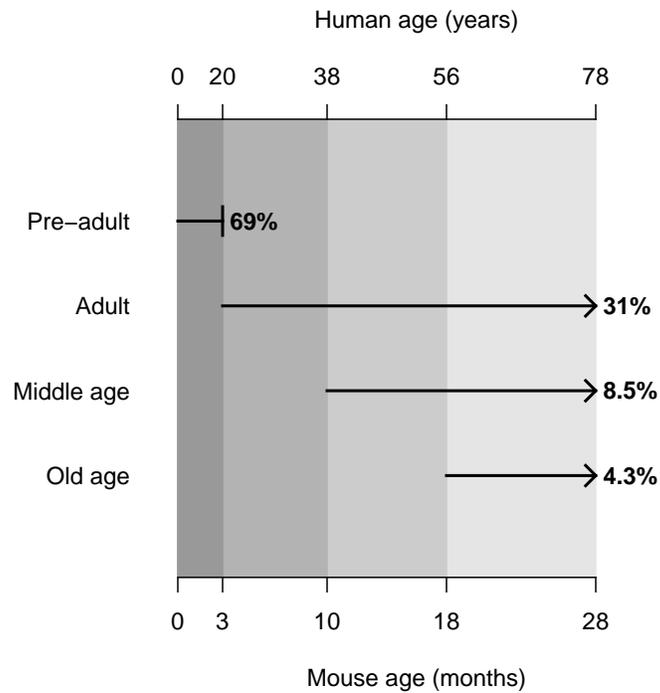}
 \caption{Percentage of cells added during various life stages. Of all the cells added to the dentate gyrus postnatally, 69\% of these will be added before adulthood, and only 31\% (95\% CI = 23.6\% to 38.1\%) of cells will be added during adulthood (69\% + 31\% = 100\%). Only 8.5\% (95\% CI = 1.0\% to 14.7\%) of new cells will be added from middle age onwards. Thus, for most of adult life, there are relatively few new cells added to the DG. The results are based on data from mice, and human ages are given as a reference. Mouse life stages and human comparisons are taken from \protect{\citet{Harrison2010}}.}
\label{fig:prolif_est}
\end{figure}

\end{document}